\documentclass[aps,prl,reprint, amsmath,amssymb,
 aps, superscriptaddress,floatfix]{revtex4-1}
\usepackage{braket}
\usepackage{graphicx}
\usepackage{hyperref}
\usepackage{comment} 
\usepackage{amsmath} 
\usepackage{amssymb}
\usepackage{wrapfig}
\usepackage{bbm}
\usepackage{url,hyperref} 
\usepackage{tikz}
\usepackage{pgfplots}
\pgfplotsset{compat=1.8}
\usetikzlibrary{pgfplots.groupplots}
\usepgfplotslibrary{external} 
\pgfplotsset{compat=newest}

\begin{document}

\newcommand{\mytitle}{The influence of the Majorana non-locality on the supercurrent}
\newcommand{\I}{\mathbbm{1}}
\newcommand{\T}{\mathcal{T}}
\newcommand{\U}{\mathcal{U}}
\newcommand{\N}{\mathcal{N}}
\newcommand{\hc}{\text{H.c.}}
\newcommand{\vlambda}{\vec{\lambda}}
\newcommand{\alphabar}{{\bar{\alpha}}}
\newcommand{\MathInclude}[1]{\medskip\fbox{\includegraphics[width=165mm]{Mathematica/#1}}\medskip}

\newcommand{\dd}{\mathrm{d}\,} 
\newcommand{\sign}{\mathrm{sign}\,}
\newcommand{\tr}{\mathrm{tr}\,}
\newcommand{\adj}{\mathrm{adj}\,}
\newcommand{\diag}{\mathrm{diag}\,} 
\newcommand{\const}{\mathrm{const}\,} 
\newcommand{\figref}[1]{\figurename~\ref{#1}} 
\newcommand{\sref}[1]{Section~\ref{#1}} 

\title{\mytitle}
\author{Alexander Schuray}
\affiliation{Institut f\"ur Mathematische Physik, Technische Universit\"at
        Braunschweig, D-38106 Braunschweig, Germany}
    \author{Alfredo Levy Yeyati}
    \affiliation{Departamento de F\'isica Te\'orica de la Materia Condensada C-V, Condensed Matter Physics Center (IFIMAC)
and Instituto Nicol\'as Cabrera, Universidad Aut\'onoma de Madrid, E-28049 Madrid, Spain}
    \author{Patrik Recher}
    \affiliation{Institut f\"ur Mathematische Physik, Technische Universit\"at
        Braunschweig, D-38106 Braunschweig, Germany}
    \affiliation{Laboratory for Emerging Nanometrology Braunschweig, D-38106 
        Braunschweig, Germany}
\date{\today}
 \begin{abstract}
We study the equilibrium Josephson current between an s-wave superconductor and a topological superconducting nanowire with Majorana bound states (MBS) at its ends. Within a low-energy model we show analytically that the non-locality of the MBS allows for a finite supercurrent to flow that otherwise would vanish. In particular, we find the critical current to be a function of the difference in the spin canting angles of the Majorana wave functions at the location of the tunnel contact. We complement our analytical calculations by numerically solving the full tight binding model and show how to extract the main features of the low-energy model from the critical current using available experimental techniques.
\end{abstract} 
\maketitle
\textit{Introduction.}---One dimensional topological superconductors (TSCs) with a $p$-wave like order parameter host Majorana bound states (MBS) at boundaries between topological trivial and non-trivial regions~\cite{Kitaev2001, Read2000, Fu2008}. These MBS are described by self-adjoint operators~\cite{Alicea2012} and have non-Abelian braiding statistics which renders them promising candidates for qubits of a topological quantum computer~\cite{Nayak2008,Alicea2011,Pachos2012}.

So far, signatures of their detection are based on electrical transport experiments~\cite{Aguado2017}, which include a zero-bias peak in the differential conductance, when an isolated MBS is tunnel contacted by a normal metallic lead~\cite{Law2009,Mourik2012,Das2012,Deng2012,Lee2013,Zhang2018,Nadj-Perge2014,Suominen2017,Nichele2017}
and the fractional Josephson effect in TSC-TSC Josephson junctions~\cite{Fu2009b,PhysRevLett.110.017003,PhysRevLett.112.077002}, which manifests itself in the missing of the odd Shapiro steps~\cite{Rokhinson2012, Dominguez2012, Virtanen2013,Wiedenmann2016}.

These key signatures of MBS rely on the assumption that MBS are spatially well separated, which is, e.g., justified in long nanowires with strong spin-orbit coupling in proximity to an $s$-wave superconductor (SOCNWs) where an applied Zeeman field can drive the system to a topological phase harboring MBS at both ends of the wire~\cite{Lutchyn2010,Oreg2010} with an exponential spatial decay of their wave functions~\cite{Klinovaja2012,Das2012}. However, recent experiments, in which a quantum dot was coupled to one end of a SOCNW, suggest that this assumption may not always hold~\cite{Deng2016,Deng2018}. Rather, these measurements can be explained theoretically by an effective model in which the quantum dot does not only couple to the closest, but also to the MBS at the other end of the wire~\cite{Schuray2017,Prada2017,Clarke2017}. Further, information on the spin canting angle of the MBS can be extracted using this Majorana non-locality~\cite{Prada2017,Deng2018}.
\begin{figure}
\centering
\centering
	\includegraphics[width=\columnwidth]{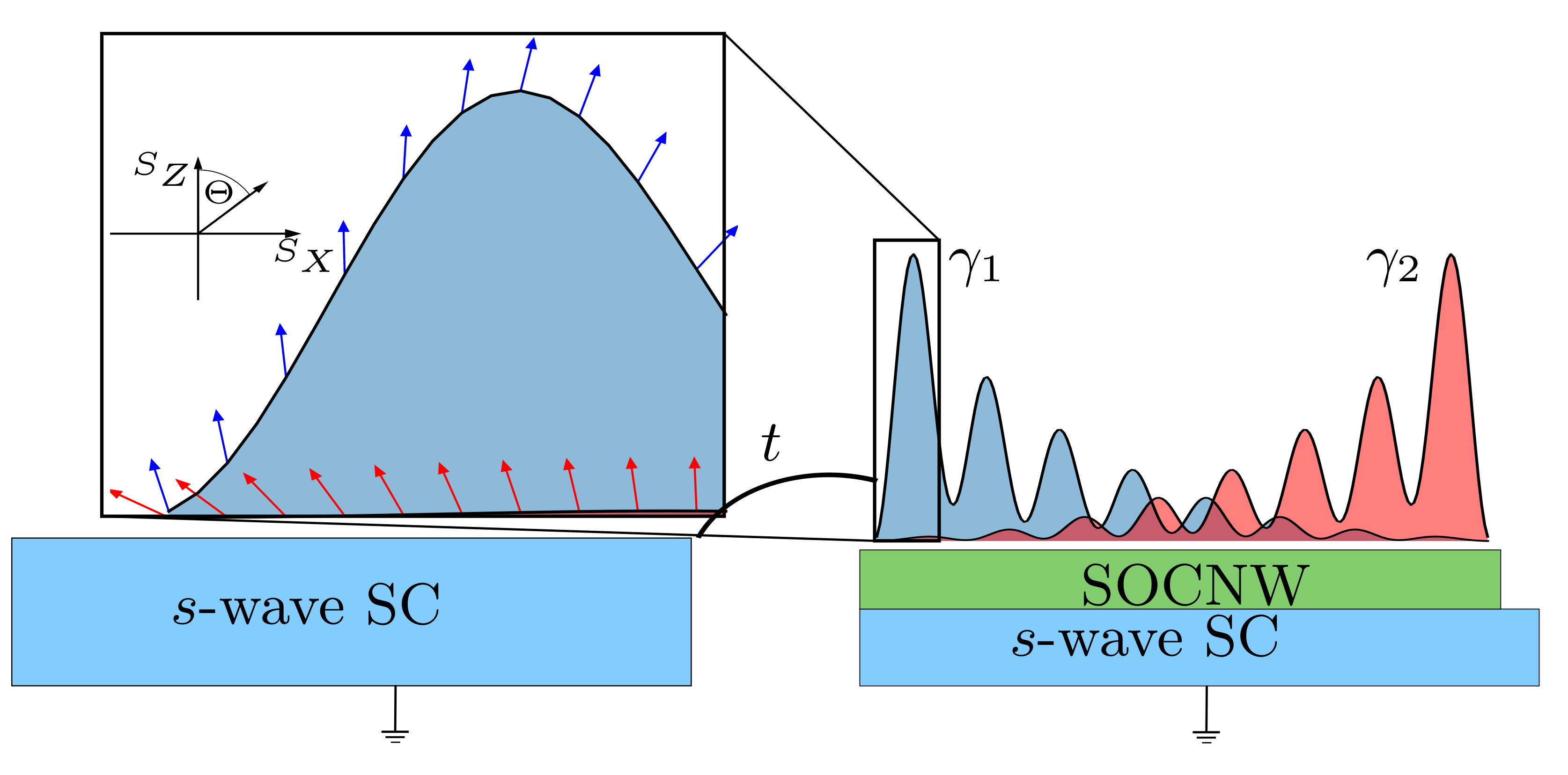}
	\caption{ Sketch of the considered Josephson junction between an s-wave superconductor and a topological superconducting nanowire including the calculated wave functions of the two MBS $\gamma_1$ and $\gamma_2$. The wave functions decay into the wire and their spin cantings (arrows) change with position along the wire. Electrons can tunnel between the $s$-wave superconducting lead and the nanowire with tunneling amplitude $t$ creating overlap with both MBS.} 
	\label{fig:Setup}
\end{figure}
Junctions consisting of a conventional $s$-wave (BCS) lead and a TSC have also been in the focus of some investigations where most of the works address on non-equilibrium transport~\cite{Peng2015,Sharma2016,Zazunov2016,Setiawan2017,Setiawan20172}. For a pure $s$-wave--$p$-wave junction it was concluded that the supercurrent is blocked~\cite{Zazunov2012}. This blockage, however, can be lifted when the BCS lead is not only coupled to one MBS but to two MBS with non-collinear spin directions~\cite{Zazunov2017,Schrade2018}. Also it was shown that BCS-SOCNW junctions exhibit a finite supercurrent which is not carried by the MBS at the interface~\cite{Ioselevich2016,Zazunov2018}.

Here, we consider a junction of a BCS lead and a SOCNW, but different to previous works we include the finite length of the SOCNW and thus the possibility to access both MBS via the Majorana non-locality (see Fig.~\ref{fig:Setup}). We calculate the equilibrium Josephson current and show that the MBS contribute a finite supercurrent in contrast to Refs.~\cite{Zazunov2012,Ioselevich2016,Zazunov2018}. Moreover, we find that the relative spin canting angles of the two MBS at the point of the tunnel contact directly govern the behavior of the critical current. These two results are the main findings of our work. To relate the supercurrent to the microscopic parameters of the junction we derive an analytically solvable low-energy model and calculate the Majorana spinor wave functions approximately.
To complete our analysis, we evaluate the Josephson current numerically using an appropriate tight binding model and show that for realistic parameters the Majorana non-locality can be extracted from the supercurrent with existing experimental techniques.

\textit{Model}.---
The total Hamiltonian of the system is described by three parts
\begin{equation}
	H=H_{BCS}+H_{NW}+H_T,
	\label{Hamtot}
\end{equation}
where $H_{BCS}$ describes the $s$-wave superconductor, $H_{NW}$ describes the proximitized nanowire and $H_T$ mediates electron tunneling between the two superconductors. The nanowire is described within the Bogoliubov-de Gennes (BdG) formalism \cite{Lutchyn2010, Oreg2010}
\begin{equation} 
	H_{NW}=\frac{1}{2}\int_0^L \Psi^\dagger(x) \mathcal{H}_{BdG}^{NW} \Psi(x) dx, 
	\label{NWBdG}
\end{equation}
where $L$ is the length of the nanowire and $H_{BdG}^{NW}$ is presented in the Nambu basis with $\Psi(x)=\left[\psi_\uparrow(x),\psi_\downarrow(x),\psi_\downarrow^\dagger(x),-\psi_\uparrow^\dagger(x)\right]^T$, and
 \begin{align}
	\mathcal{H}_{BdG}^{NW}=&\left[\left(-\frac{\hbar^2}{2m^*}\partial_x^2 -\mu\right)-i \alpha \partial_x \sigma_y\right]\tau_z\notag\\
	&+V_Z\sigma_z+\Delta\tau_x.
	\label{NW}
\end{align}
Here, $m^*$ is the effective electron mass, $\mu$ is the chemical potential, $\alpha$ is the Rashba parameter, $V_Z$ is the Zeeman energy and $\Delta$ is the induced $s$-wave pairing. The Pauli matrices $\sigma_i$ and $\tau_i$ act in the spin and particle-hole space, respectively. The topological non-trivial phase with emerging MBS is present for $V_{Z}>\sqrt{\Delta^2+\mu^2}$~\cite{Lutchyn2010,Oreg2010}.
The BCS lead is modeled by
\begin{align}
	H_{BCS}=&\sum_{\mathbf{k}\sigma}\xi_k c_{\mathbf{k}\sigma}^{\dagger}c_{\mathbf{k}\sigma}\notag\\
	&+\sum_\mathbf{k} \Delta_{BCS} (c_{\mathbf{k}\uparrow}^\dagger c_{-\mathbf{k}\downarrow}^\dagger+c_{-\mathbf{k}\downarrow}c_{\mathbf{k}\uparrow}),
	\label{HBCS}
\end{align}
where $\Delta_{BCS}$ is the superconducting pairing in the BCS lead and $\xi_k=\varepsilon_k-\mu$ is the single particle energy in the normal phase~\cite{Tinkham2004}.
We choose a gauge in which the phase difference between the two superconductors only appears in the tunneling Hamiltonian
 \begin{equation} 
	 H_T=\sum_{\mathbf{k}\sigma} te^{i\frac{\varphi}{2}} c_{\mathbf{k}\sigma}^\dagger \psi_{\sigma}(0)+h.c.,
	\label{HT}
\end{equation}
with a momentum and spin independent tunneling amplitude $t$.

In order to calculate the ground-state energy analytically we treat the tunneling Hamiltonian as a perturbation. $H_{NW}$ and $H_{BCS}$ can be diagonalized with canonical transformations
$\psi_\sigma(x)=\sum_n\alpha_{n\sigma}(x)\eta_n+\beta_{n\sigma}^*(x)\eta_n^\dagger$ and $c_{\mathbf{k}\sigma}=u_k \gamma_{\mathbf{k}\sigma}+\sigma v_k\gamma_{-\mathbf{k}\overline{\sigma}}^\dagger$. Here, $u_k$ and $v_k$ are the standard BCS coherence factors~\cite{Tinkham2004} and $\alpha_{n\sigma}(x)$ and $\beta_{n\sigma}(x)$ are the wave functions of the corresponding Bogoliubov quasiparticles, $n$ corresponds to the positive eigenenergies $\varepsilon_n>0$ of $H_{BdG}^{NW}$, and with $\eta_n^\dagger$ and $\gamma_{\mathbf{k}i}^\dagger$ creating an excitation in the SOCNW and the BCS lead, respectively such that $\gamma_{\mathbf{k}\sigma}\ket{0}=\eta_n\ket{0}=0$, where $\ket{0}=\ket{0}_{BCS}\otimes\ket{0}_{NW}$ is the unperturbed ground state with energy $E_0$. To second oder in the tunneling Hamiltonian the ground state energy can be written as $E_{0}^{(2)}=E_0+E_{\varphi}^{(2)}$,  with the phase dependent correction
  \begin{align}
	 E_\varphi^{(2)}=&\sum_ne^{i\varphi}t^2\left(\alpha_{n\uparrow}(0)\beta_{n\downarrow}^*(0)-\alpha_{n\downarrow}(0)\beta_{n\uparrow}^*(0)\right)\notag\\
	 &\times f\left(\frac{\varepsilon_n}{\Delta_{BCS}}\right)+h.c.,
	 \label{phicorrection}
 \end{align}  
 with $f(x)=\nu(0)\int_{0}^{y_{max}}dy(\sqrt{1+y^2}(\sqrt{1+y^2}+x))^{-1}$,
 where $\nu(0)$ is the density of states in the BCS lead at the Fermi level and $y_{max}=\hbar\omega_D/\Delta_{BCS}$ with $\omega_D$ the Debye frequency. The wave function contributions in Eq.~(\ref{phicorrection}) take the form of a singlet, reflecting the $s$-wave pairing in the BCS lead. It also explains the proposed blockage of the supercurrent in a pure $s$-wave-$p$-wave junction~\cite{Zazunov2012}, because for a pure p-wave superconductor
 \begin{equation}
	 \alpha_{n\uparrow}(0)\beta_{n\downarrow}^*(0)-\alpha_{n\downarrow}(0)\beta_{n\uparrow}^*(0)=0\quad\forall n.
	 \label{Zazunovconnection}
 \end{equation}
 So in general, contributions to the Josephson current originate from residual $s$-wave pairing in the SOCNW~\cite{Zazunov2018}.  

 \textit{Josephson current in the low-energy model.}---
In the topologically non-trivial phase of the SOCNW the low energy physics is governed by the two MBS described by Hermitian operators $\gamma_1$ and $\gamma_2$ satisfying $\{\gamma_i,\gamma_j\}=2\delta_{ij}$.
Using only the two MBS the Hamiltonian for the SOCNW reduces to
\begin{equation} 
	H_{NW}=i\varepsilon\gamma_1\gamma_2,
	\label{effectiveNW}
\end{equation}
where, following Ref.~\cite{DasSarma2012}, $\varepsilon$ can be expressed with the microscopic parameters used in Eq.~(\ref{NW})~\footnote{Here, $\varepsilon\approx\hbar^2k_{F,\text{eff}}\frac{e^{-2L/\xi}}{m^*\xi}\cos\left(k_{F,\text{eff}} L\right)$, where the wave number $k_{F,\text{eff}}$ and the Majorana localization length $\xi$ are functions of the microscopic parameters of $H_{NW}$. In App. A, the derivation of both parameters is explained in detail.}. This low-energy model the Hamiltonian is diagonal in the basis of the non-local fermion level $\eta_1=(\gamma_1+i\gamma_2)/2$  which can be unoccupied (even parity) or occupied (odd parity).
\begin{figure}
\centering
	
\includegraphics[width=\columnwidth]{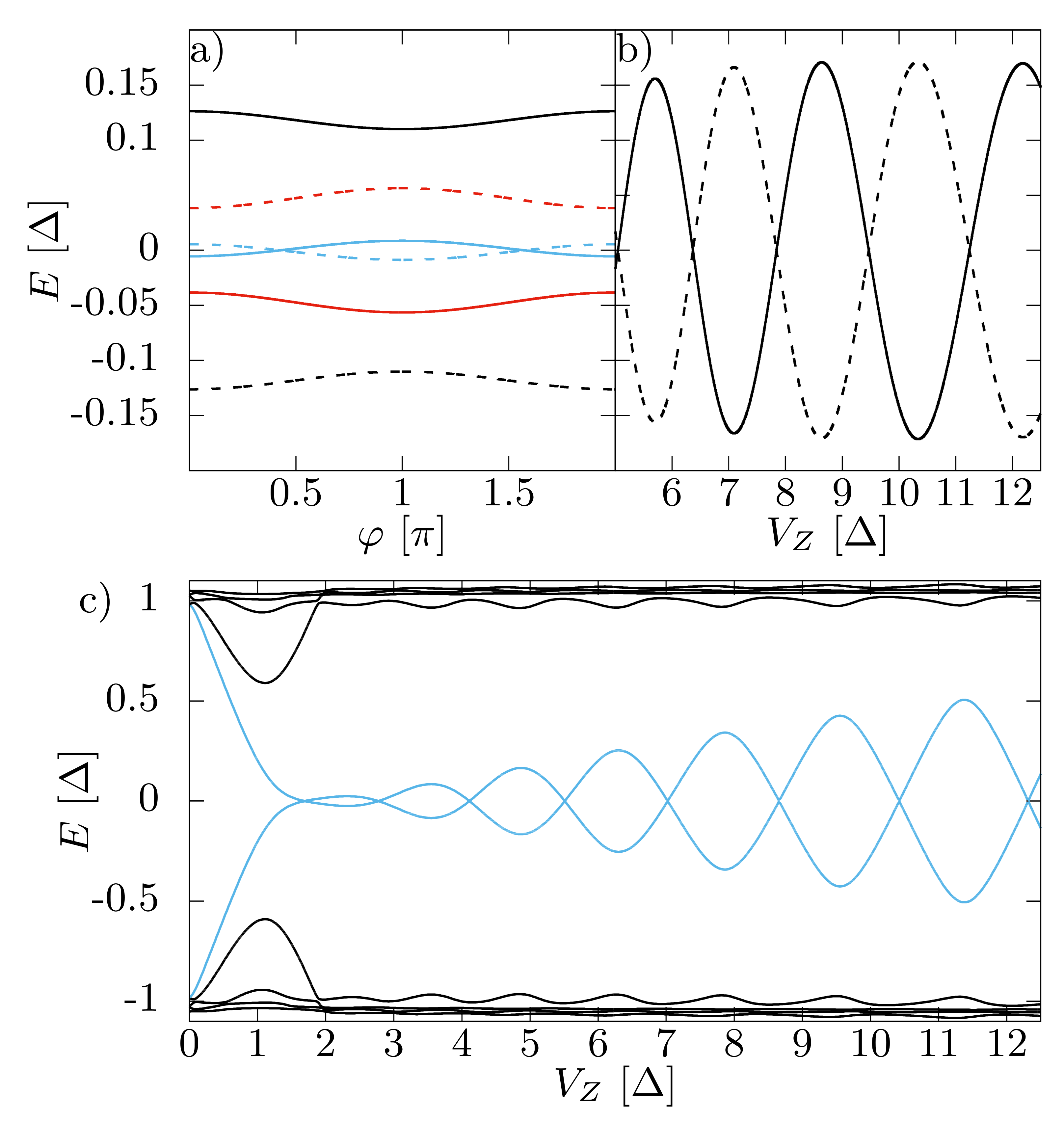}
\caption{a) and b): Effective low energy spectrum (Eq.~(\ref{HEO})) as function of the superconducting phase difference for different applied Zeeman fields (blue: $V_Z=5.05\Delta$, black: $V_Z=6.0\Delta$, red: $V_Z=6.5\Delta$) (a)) and as function of the applied Zeeman field for $\varphi=\pi/2$ (b)). Dashed lines correspond to odd parity states, while full lines correspond to even parity states. c): Corresponding BdG spectrum of the tight binding Hamiltonian $H$ with $t_S=10\Delta$, $\tilde{t}=2.96$ meV and $\Delta_{BCS}=\Delta$. The lowest energy levels (blue) correspond to the hybridized MBS after the topological phase transition.  The other microscopic parameters are $m^*=0.015m_e$, $\Delta=0.2$ meV, $\mu=0$, $\alpha=20$ meV nm, $\Gamma=0.004$ meV, $L=1.3$ $\mu$m and $N=M=100$.}
\label{fig:effspec}
\end{figure}  
In this approximation the electron annihilation operator at the tunnel junction
$\psi_\sigma(0)\approx\Lambda_{1\sigma}(0)\gamma_1+\Lambda_{2\sigma}(0)\gamma_2$,
where $\Lambda_{n\sigma}(x)$ are the electron-components of the Majorana wave functions.
In the limit $\varepsilon\ll\Delta_{BCS}$ and $\omega_{D}\rightarrow\infty$ we find for the eigenenergies within second order degenerate perturbation theory in $H_T$
\begin{align}
	\label{Heffeo}
E_{o(e)}^{(2)}=&\mp\varepsilon\pm 2\pi \nu(0)\\
	&\times\left(it^2\left[\Lambda_{1\uparrow}(0)\Lambda_{2\downarrow}(0)-\Lambda_{1\downarrow}(0)\Lambda_{2\uparrow}(0)\right]e^{i\varphi}+h.c.\right),\notag
\end{align} 
where the lower signs correspond to the odd and the upper ones to the even parity states of the junction. 
The spinor wave functions for the MBS can be calculated analytically when we employ the approximation that both MBS reside in semi-infinite wires, which are then cut off at a given length $L$.
A detailed calculation for these wave functions can be found in App.~\ref{apA}. As the Majorana wave functions have no spin component in the y-direction~\cite{Sticlet2012}, we write
\begin{align} 
	\begin{pmatrix}
		\Lambda_{n\uparrow}(0)\\
		\Lambda_{n\downarrow}(0)
	\end{pmatrix}=i^{n-1}\kappa_n\left(\begin{matrix}
			\cos(\Theta_n/2)\\
			\sin(\Theta_n/2)
	\end{matrix}\right),
\end{align}  
where $\kappa_n$ is real valued. We approximate $\kappa_2=\kappa_1e^{-L/\xi}$, where $\xi$ is the localization length of the MBS, because previous works have shown that the MBS are exponentially localized~\cite{DasSarma2012,Klinovaja2012}. The spin canting angle $\Theta_n$ at position $x=0$ can differ for the two MBS. 
Inserting this parametrization into Eq.~(\ref{Heffeo}) we find 
 \begin{equation}
	E _{e(o)}^{(2)}=\mp\varepsilon\pm\Gamma\cos(\varphi)\sin\left(\frac{\Theta_1-\Theta_2}{2}\right)e^{-L/\xi},
	 \label{HEO}
 \end{equation}
 with $\Gamma=4\pi\kappa_1^2t^2\nu(0)$ where we assumed that $t$ is real. The two parities are distinguished by a zero and a $\pi$-junction behavior, respectively (see Fig.~\ref{fig:effspec}a)). The MBS hybridization energy $\varepsilon$ oscillates as a function of $V_Z$ and its amplitude is rising, because $\xi$ grows with increasing $V_Z$ (see Fig.~\ref{fig:effspec}b)). Moreover, Eq.~(\ref{HEO}) shows that when the spin canting of the two MBS is the same the Josephson current will be blocked, conversely, the Josephson current will be maximal when the spins point in opposite directions. Due to the localization of the MBS the Josephson current will be exponentially suppressed if $L$ exceeds $\xi$. 
 
The equilibrium ground state Josephson current is obtained by taking the derivative of the ground state energy with respect to the phase
\begin{align}
	I(\varphi)&=\frac{2e}{\hbar}\partial_\varphi\text{min}(E_{e}^{(2)},E_{o}^{(2)})\notag\\
	&=I_C\text{ sgn}\left(E_{e}^{(2)}(\varphi)-E_{o}^{(2)}(\varphi)\right)\sin(\varphi).
	\label{CPR}
\end{align} 
\begin{figure} 
\centering
	
\includegraphics[width=\columnwidth]{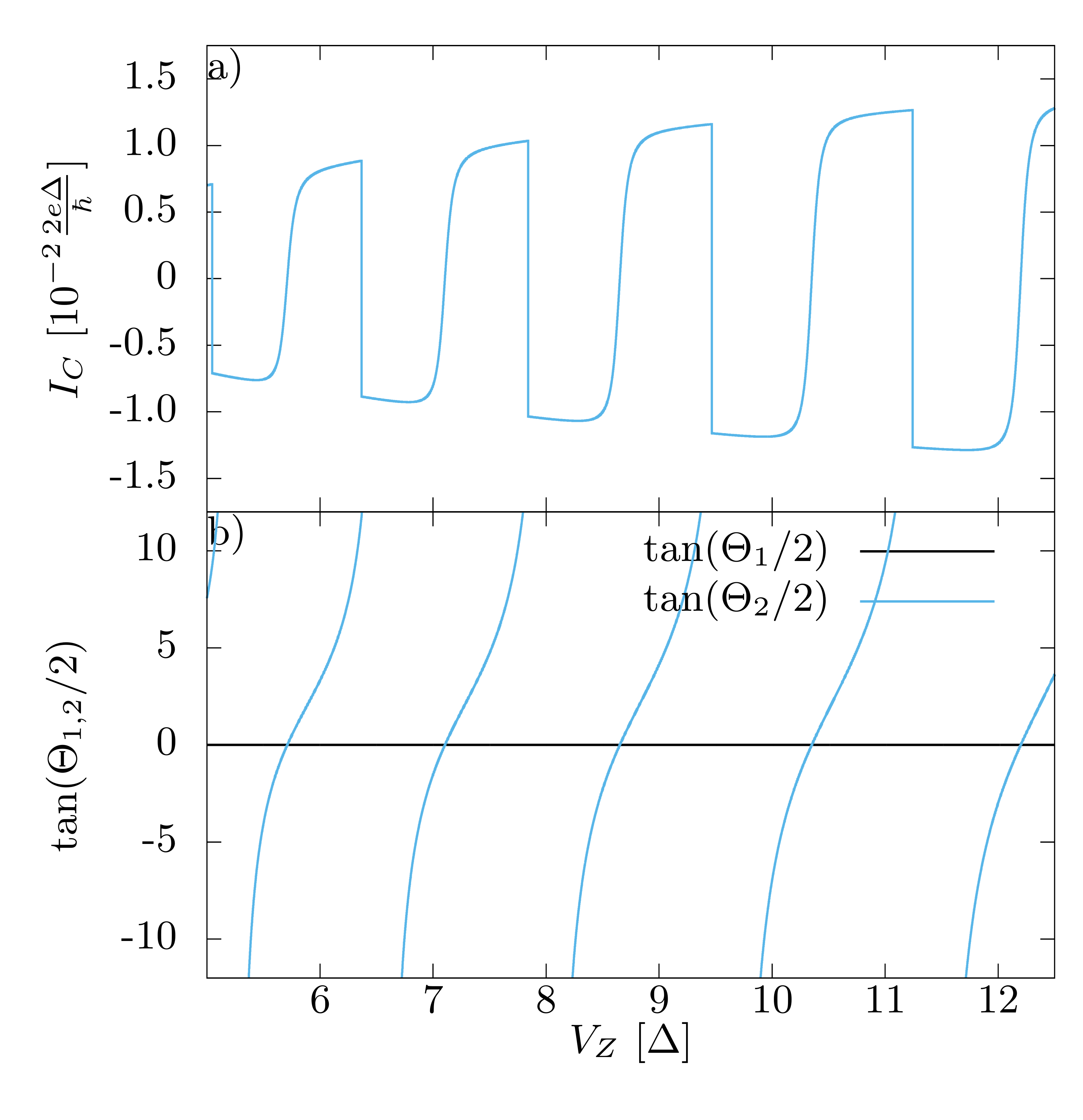}
\caption{a) Critical current in the ground state of the low-energy model and b) tangent of the spin canting angles at position $x=0$ of the two MBS (1,2) as a function of Zeeman field $V_Z$. The other parameters are the same as in Fig.~\ref{fig:effspec}. The jumps in the supercurrent occur when there is a parity switching of the ground state, while the oscillatory part is due to spin canting rotation of the MBS. When the difference of the spin canting angles is zero the critical current vanishes.}
	\label{fig:effectiveI}
\end{figure}
In addition to informations on the Majorana spinor rotation with Zeeman field, also the parity changes in the ground state are visible in the supercurrent. As seen in Fig.~\ref{fig:effectiveI}a) the critical current $I_C$ jumps and switches sign whenever the ground state parity changes sign. In accordance with previous works~\cite{Sticlet2012,Prada2017} the MBS at their respective ends are nearly polarized along the applied magnetic field direction. The rotation of the spin difference is mainly governed by the more distant MBS (Fig.~\ref{fig:effectiveI}b)). Its spin rotates along the length of the wire due to the spin orbit coupling. The magnitude of the critical current rises as the applied Zeeman field is increased, because the Majorana localization length $\xi$ is increased.

\begin{figure} 
\centering
	\includegraphics[width=\columnwidth]{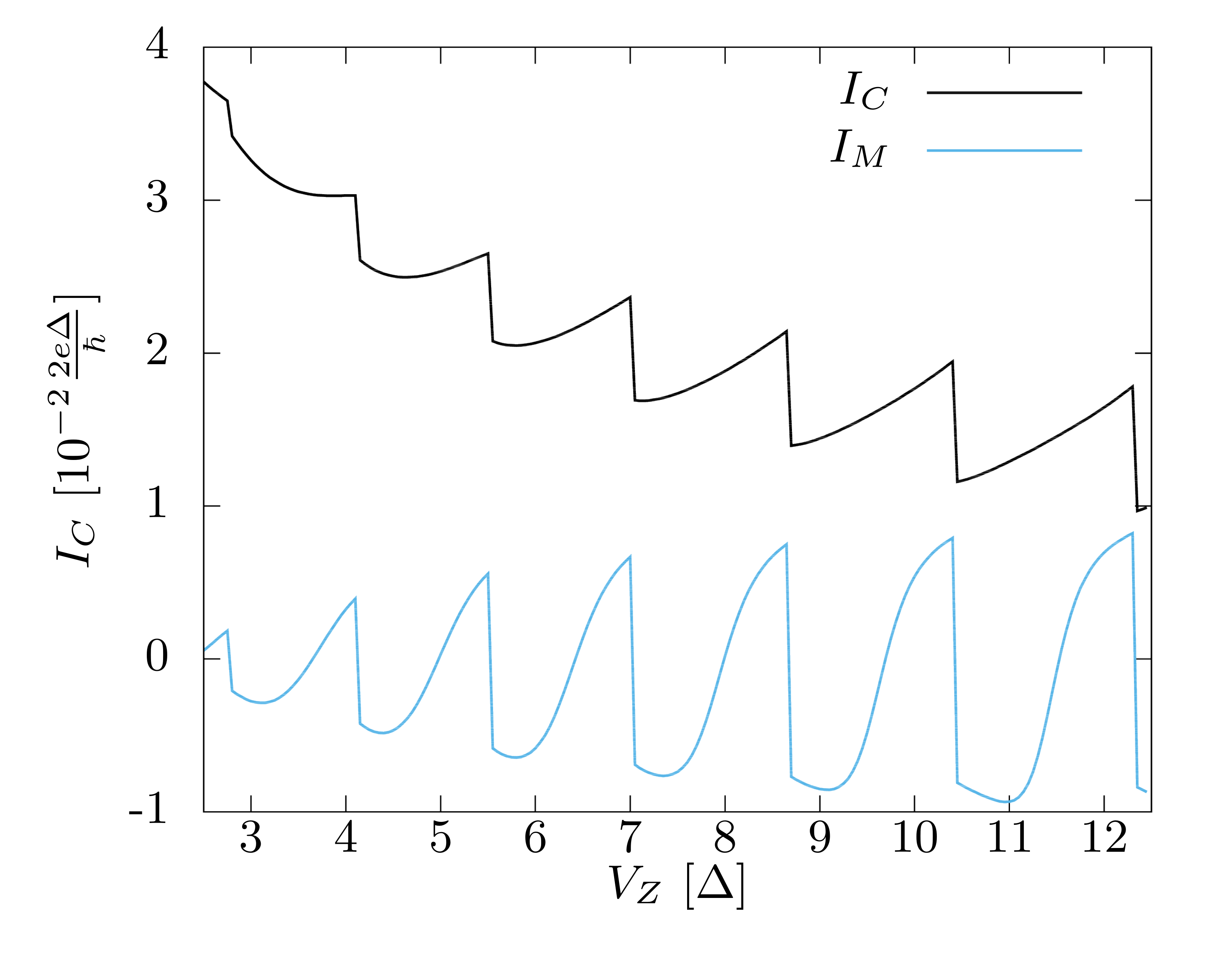}
	\caption{Numerically calculated ground state critical current $I_C$ (black) and Majorana contribution to the critical current $I_M$  (blue) extracted following the scheme proposed in the main text. Jumps in the critical current occur at parity crossings in the spectrum. Parameters are the same as in Fig.~\ref{fig:effspec} c). }
	\label{fig:tightbinding1}
\end{figure}

\textit{Tight binding analysis.}---We contrast the low-energy model with a tight binding approach by discretizing the Hamiltonian $H$ and using kwant~\cite{Groth2014}.
The details of the tight binding calculation can be found in App.~\ref{app:numham} and \ref{sec:wireana}. In order to calculate the Josephson current we diagonalize the resulting Hamiltonian numerically. In Fig.~\ref{fig:effspec}c) the oscillations of the hybridization energy of the MBS as a function of $V_Z$ can be seen after the topological phase transition consistent with~\cite{DasSarma2012}. The ground state Josephson current is then calculated as
\begin{equation}
	I(\varphi)=\frac{2e}{h}\partial_\varphi\sum_{E_i<0}E_i(\varphi),
\end{equation}
and the critical current $I_C$ is given as
\begin{equation}
	I_C=\max_\varphi I(\varphi),
	\label{eq:IC}
\end{equation}
which is shown in Fig.~\ref{fig:tightbinding1}.	
The critical current in the low-energy model (see Fig.~\ref{fig:effectiveI}a)) and in the tight-binding model obviously share the common features of jumps at parity crossings and an oscillating behavior between them. We attribute the latter effect to the rotation of the spin canting angle of the more distant MBS. That the jumps are associated to parity crossings is clear from the analytic solution Eq.~(\ref{HEO}) and is further substantiated by corresponding sign changes of the Majorana charge defined in Ref.~\cite{Escribano2017} (see App.~\ref{sec:wireana}). However, we find that the contributions of the higher energy Andreev bound states, neglected in the low-energy model, also contribute to the ground state critical current due to residual $s$-wave pairing in the higher energy excitations, even after the topological phase transition~\cite{Zazunov2018}. These additional contributions disguise the oscillatory behavior of $I_C$ contributed by the MBS showing vanishing critical currents for certain values of $V_Z$ in the low-energy model.

To experimentally extract the Majorana contribution $I_M$ to $I_C$, we propose to measure $I_C$ for a certain set of parameters $\Delta_Z$. For fixed $\Delta_Z$, $I_C$ is measured for the two particle parities distinguishing the two low-energy states spanned by the two MBS. The switching of $I_C$ happens within the quasiparticle poisoning time $T_P\approx100\mu$s~\cite{OFarrell2018} and its difference subtracts the common background contributions from the high energy states. Moreover, due to particle-hole symmetry, the difference between the two values for $I_C$ will bring out twice the desired Majorana contribution $I_M$ (shown in Fig.~\ref{fig:tightbinding1})~\footnote{The parameters used in Fig.~2c) and Fig.~4 for the tight-binding model correspond to the same tunneling rate $\Gamma=0.004$ meV as used in Figs.~2a), 2b) and 3.}. The suggested experiment is feasible as currents with a sensitivity of $10^{-2}2e\hbar/\Delta$ can be measured on a time scale of 10 $\mu$s~\footnote{Private communication with Cristian Urbina.}. Another proposal to reveal the Majorana contributions in the critical current is discussed in App.~\ref{sec:adiabatic}.

In summary, we studied a Josephson junction consisting of a standard BCS $s$-wave superconductor and a topological superconducting nanowire. Going beyond existing considerations, we analyzed in detail the role of the distant MBS in the critical current of such a topological junction. 
We found analytically that the size of the supercurrent carried by the low-energy MBS is directly proportional to the spin-singlet component of their wave function overlap at the location of the junction. This overlap depends in particular on the spin canting angle of the distant Majorana in an oscillatory fashion which could be probed by changing the Zeeman field. Considering the whole spectrum of the nanowire numerically in a tight-binding approach, we found that the residual s-wave pairing of the high energy states contribute a background to the critical current, that, however, could be neutralized by measuring and comparing the critical current in the different parity states. 

\textit{Note added.}-- While finishing the manuscript, the preprint \cite{Cayao2018} appeared where the Josephson current between a trivial- and a topological superconductor nanowire of finite size is studied. The paper contains a purely numerical analysis which does not focus on the Majorana non-locality. 

We thank Cristian Urbina, Reinhold Egger and Miguel Alvarado for useful discussions and suggestions and the Lower Saxony PhD-programme “Contacts in Nanosystems”, the Research Training Group GrK1952/1 “Metrology for Complex
Nanosystems”, the Braunschweig International Graduate School of Metrology B-IGSM, the Spanish MINECO through Grants No.~FIS2014-55486-P, FIS2017-84860-R and the "Mar\'{\i}a de Maeztu" Programme for Units if Excellence in R\&D (MDM-2014-0377) for support.

\appendix
\section{Majorana wave functions} 
\label{apA}

In the following, we want to calculate the spinor components for both MBS of a SOCNWs in order to relate the Josephson current to microscopic parameters. For the calculations we closely follow~\cite{Prada2017}.
To start with, we consider a spin orbit coupled nanowire in proximity to a superconductor and an applied Zeeman field. Its BdG Hamiltonian in the Nambu basis is given as
\begin{align} 
	&H_\text{BdG}=\\
	&\left(\begin{smallmatrix}
         -\frac{\hbar^2}{2m^*}\partial_x^2-\mu+V_Z&-\alpha\partial_x&\Delta&0\\
         \alpha\partial_x&-\frac{\hbar^2}{2m^*}\partial_x^2-\mu-V_Z &0&\Delta \\
 	\Delta&0&\frac{\hbar^2}{2m^*}\partial_x^2+\mu+V_Z&\alpha\partial_x\\
        0& \Delta &-\alpha\partial_x&\frac{\hbar^2}{2m^*}\partial_x^2+\mu-V_Z\notag
        \end{smallmatrix}
\right)
\end{align}
As we search for Majorana like solutions the hole-like and electron-like components of its wave function
\begin{align}
&	\alpha^{(L,R)}(x)\\
	&=\left(u_{\uparrow}^{(L,R)}(x),u_\downarrow^{(L,R)}(x),(v^{(L,R)}_\downarrow)^* (x),-(v^{(L,R)}_\uparrow)^* (x)\right)^T,\notag
	\label{defwavefunc}
\end{align}
have to satisfy
\begin{equation}
 (v_{\downarrow\uparrow}^{(L,R)})^*=\lambda u_{\downarrow\uparrow}^{(L,R)},
\end{equation}
where $\lambda=\pm1$ for the left (L) and right (R) MBS and need to be zero energy solutions. Here, we denote the MBS with left and right for clarity. This reduces the four dimensional eigenvalue problem to a two dimensional problem
\begin{equation}
 \left(\begin{matrix}
  \frac{-\hbar^2}{2m}\partial_x^2-\mu+V_Z&-\alpha\partial_x+\lambda\Delta\\
  \alpha\partial_x-\lambda\Delta& \frac{-\hbar^2}{2m}\partial_x^2-\mu-V_Z
 \end{matrix}\right)\left(\begin{matrix}
 u_{\uparrow}^{(L,R)}(x)\\
 u_{\downarrow}^{(L,R)}(x)
 \end{matrix}\right)=0.
 \label{HM}
\end{equation}
We are now considering two kinds of solutions, a solution for the left MBS which decays exponentially for $x>0$ and a solution for the right MBS at position $L$ which exponentially
decays in the other direction. However, the solutions of Eq.~(\ref{HM}) which also satisfy the boundary conditions $u_{\downarrow\uparrow}^{(L,R)}(0)=u_{\downarrow\uparrow}^{(L,R)}(L)=0$ do not exist.
So we consider two independent semi infinite nanowires which range from $x=0$ to $x=\infty$ for the left MBS and from $x=0$ to $x=-\infty$ for the right MBS which leads to the boundary conditions
\begin{align}
 u^L_\sigma(0)&=u^L_\sigma(\infty)=0\notag\\
 u^R_\sigma(0)&=u^R_\sigma(-\infty)=0.
 \label{boundary}
\end{align}
We use the ansatz 
\begin{equation}
 \left(\begin{matrix}
 u_{\uparrow}^{(L,R)}(x)\\
 u_{\downarrow}^{(L,R)}(x)
 \end{matrix}\right)\propto\left(\begin{matrix}
 u_{\uparrow}^{(L,R)}\\
 u_{\downarrow}^{(L,R)}
 \end{matrix}\right)e^{a x},
\end{equation}
which leads to
\begin{align}
\label{a}
&\left(\frac{\hbar^2}{2m}\right)^2 (a^{(L,R)})^4+\left(\alpha^2+\mu\frac{\hbar}{2m}\right) (a^{(L,R)})^2\\
&+2\lambda\alpha\Delta a^{(L,R)} + \mu^2+\Delta^2-B^2=0.\notag
\end{align} 
For the anticipated decay, we need $\text{Re}[a^L]<0$ and $\text{Re}[a^R]>0$. For the spinor components we find
\begin{equation}
 \left(\begin{matrix}
 u_{\uparrow}^{(L,R)}\\
 u_{\downarrow}^{(L,R)}
 \end{matrix}\right)\propto\left(\begin{matrix}
 \frac{\hbar^2}{2m}(a^{(L,R)})^2+B+\mu \\
 a^{(L,R)}_i \alpha - \lambda\Delta
 \end{matrix}\right).
\end{equation} 
For $B^2-\Delta^2-\mu^2>0$, so in the topologically non trivial regime, we find 3 solutions of Eq.~(\ref{a}) for both MBS which satisfy the restraints to their real parts.
They can be parametrized as $a^{(L,R)}_{1/2}=c^{(L,R)}_1\pm i c^{(L,R)}_2$ and 
$a^{(L,R)}_{3}=-c^{(L,R)}_1+\sqrt{(c^{(L,R)}_1)^2+4(B^2-\Delta^2-\mu^2)/((c^{(L,R)}_1)^2+(c^{(L,R)}_2)^2)}$, where $c^{(L,R)}_1$ and $c^{(L,R)}_2$ are real valued.
The wave functions of the MBS can then be written as
\begin{align}
 \Psi_{(L,R)}(x)&=\left(\begin{matrix}
 u_{\uparrow}^{(L,R)}(x) \\
 u_{\downarrow}^{(L,R)}(x)
 \end{matrix}\right)\\
 &=\sum_{i=1}^3C^{(L,R)}_i\left(\begin{matrix}
 \frac{\hbar^2}{2m}(a^{(L,R)}_i)^2+B+\mu \\
 a^{(L,R)}_i \alpha - \lambda\Delta
 \end{matrix}\right)e^{a^{(L,R)}_i x}\notag
\end{align}
Here, the factors $C^{(L,R)}_i$ follow from the boundary conditions Eqs.~(\ref{boundary}) (4 equations: $\uparrow$, $\downarrow$, $R$, $L$) and normalization (2 equations: $L$, $R$). To calculate the wave function at $x=0$ we neglect the solution corresponding to $a_{3}^R$ for the right MBS, because $|a_{3}^R|$ is larger than $|\text{Re}(a_{1/2}^{(L,R)})|$. The real and imaginary part of $a_{1}^{(L,R)}$ then correspond to the Majorana localization length $\xi$ and the wave number $k_{F,\text{eff}}$. To extract the spin canting angle of the MBS at $x=0$ we consider
\begin{equation} 
	\lim_{x\rightarrow 0}\frac{u_{\downarrow}^{L}(x)}{u_{\uparrow}^{L}(x)}=\tan\left(\frac{\Theta_{1}}{2}\right),\quad \frac{u_{\downarrow}^{R}(-L)}{u_{\uparrow}^{R}(-L)}=\tan\left(\frac{\Theta_{2}}{2}\right).	
	\label{eq:tan}
\end{equation}
The arctangent then reveals the spin canting angles of the MBS.
\section{Discretization of the Hamiltonian}
\label{app:numham}
To discretize the model Hamiltonian we use the finite differences method. The discretized Hamiltonian reads for the SOCNW
\begin{align}
	H_\text{W}=\sum_{j=1}^N& \Psi_j^\dagger \left[(\frac{\hbar^2}{m^*a^2}-\mu)\tau_z+V_Z\sigma_z+\Delta\tau_x\right]\Psi_j\notag\\
	&\Psi_j^\dagger \left[-\frac{\hbar^2}{2m^*a^2}\tau_z+i\frac{\alpha}{a}\tau_z\sigma_y\right]\Psi_{j-1}+h.c.,
\end{align}
where $a=L/N$ and $\Psi_j^\dagger$ is the four component creation operator in the Nambu basis as before.
The tight binding Hamiltonian for the $s$-wave lead with $M$ sites is
\begin{equation}
	H_{\text{BSC}}=\sum_{j=1}^M\left\{t_{S} \sum_\sigma c_{j,\sigma}^\dagger c_{j-1,\sigma}+\Delta_{BCS}e^{i\varphi}c_{j,\uparrow}c_{j,\downarrow}+h.c.\right\},
\end{equation}
where the hopping energy $t_\text{S}$ is connected to the bandwidth of the superconductor and we choose the chemical potential to be in the middle of the band. We consider a tunnel coupling between the first site of the SOCNW and the last site of the SC lead
 \begin{equation}
	 H_\text{T}=\tilde{t} c_{M,\sigma}^\dagger\psi_{1,\sigma}+h.c.
\end{equation}

\section{Characterization of spectral and charge properties of the nanowire}
\label{sec:wireana}
Here, we want to discuss the properties of the SOCNW in more detail. 
In general, one would think that the high energy contributions to the Josephson current should be smaller than those of the in-gap states (here the MBS contribution), because of the suppression factor $f(\frac{\varepsilon_n}{\Delta_{BCS}})$ in Eq.~(\ref{phicorrection}). But due to the Majorana localization the low energy contribution is exponentially suppressed with the length of the wire, while the suppression of the extended states above the gap is not that strong. By fine tuning the parameters (very short wires, small BCS gap) it is possible to enter a regime in which the higher energy contributions to the Josephson current are more suppressed than those of the two MBS. However, in this regime the low-energy model derived before looses its validity.\\
\begin{figure}
\centering
	
	\includegraphics[width=\columnwidth]{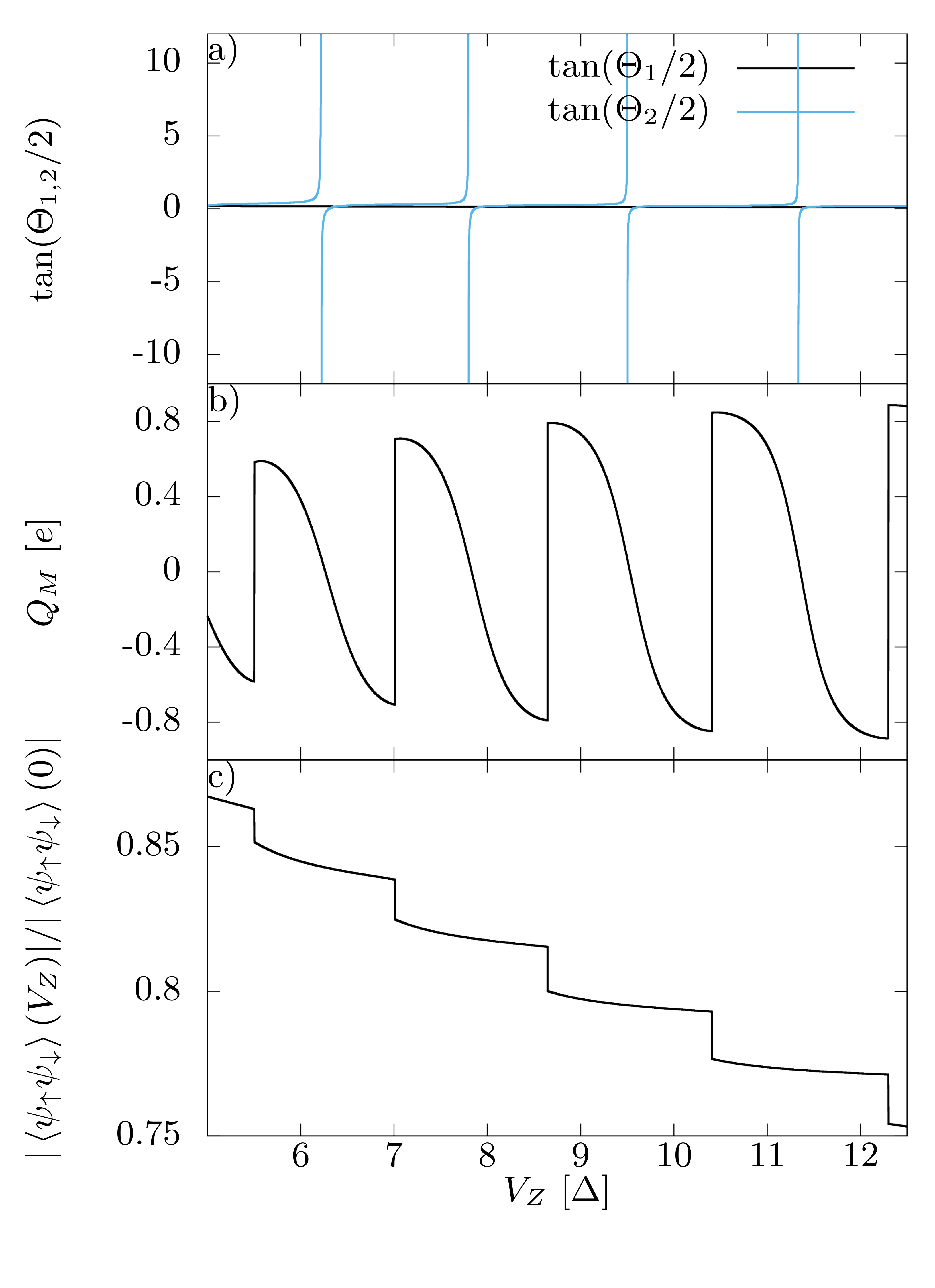}
\caption{a) Tangent of the spin canting angles, b) Majorana charge and c) $s$-wave pairing amplitude at $x=0$ as function of Zeeman field for a finite size Majorana wire with length $L=1.3$ $\mu$m. The other parameters are as in Fig.~\ref{fig:tightbinding1}. The spin canting angle $\Theta_2$ of the right Majorana shows an oscillatory behavior as a function of the Zeeman field, while $\Theta_1$ remains nearly constant. The Majorana charge and the $s$-wave pairing amplitude exhibit jumps at parity crossings.}
	\label{fig:wireanalysis}
\end{figure}    
The residual $s$-wave pairing at the end of the wire can be calculated using $|\braket{\psi_{\uparrow}(0)\psi_{\downarrow}(0)}|=|\sum_n\alpha_{n\uparrow}(0)\beta_{n\downarrow}^*(0)-\alpha_{n\downarrow}(0)\beta_{n\uparrow}^*(0)|$ from tight binding simulation. As shown in Fig.~\ref{fig:wireanalysis}c), the pairing amplitude is decreasing with increasing Zeeman field and shows jumps at parity crossings so that these jumps can be associated with the contribution of the overlapping MBS. These contributions are much smaller than those from the higher energy excitations.\\
The Majorana wave functions can be obtained numerically~\cite{Escribano2017} via $\gamma_1=\eta_{+1}+\eta_{-1}$ and $\gamma_2=-i(\eta_{+1}-\eta_{-1})$, where $\eta_{\pm1}$ are the two energy eigenstates closest to zero. Using the electron parts of these spinor wave functions we can calculate the spin canting angles of the two MBS at the left  end of the wire. For large Zeeman fields $V_Z\gg\Delta$ the spin canting of the first MBS is nearly constant at $\Theta_1\approx 0$, while the second spin canting angle oscillates as a function of $V_Z$. This is in agreement with our low-energy model calculations. However, there are quantitative differences as seen in Fig.~\ref{fig:wireanalysis}a) which we attribute to the simplifications we made in order to find the analytic results for the Majorana spinors.\\
We also consider the so called Majorana charge~\cite{Escribano2017}
\begin{equation} 
	Q_M=e\int_0^Ldx\sum_\sigma u_{\sigma}^L(x)u_{\sigma}^R(x),
	\label{MC}
\end{equation}
where $u^{(L,R)}$ are the electron components of the left (right) Majorana wave function. The abrupt sign changes in the Majorana charge (displayed in Fig.~\ref{fig:wireanalysis}b)) we attribute to parity changes of the ground state. At these points, the absolute value of $Q_M$ is maximal consistent with Ref.~\cite{Escribano2017}, where it was shown that the absolute value of the Majorana charge is highest at the parity crossings. The positions of the jumps in $Q_M$ indeed coincide with the positions of the jumps in the critical current $I_C$ in the main text (Fig.~4).
\section{Adiabatic switching}
\label{sec:adiabatic} 
\begin{figure} 
\centering
	\includegraphics[width=\columnwidth]{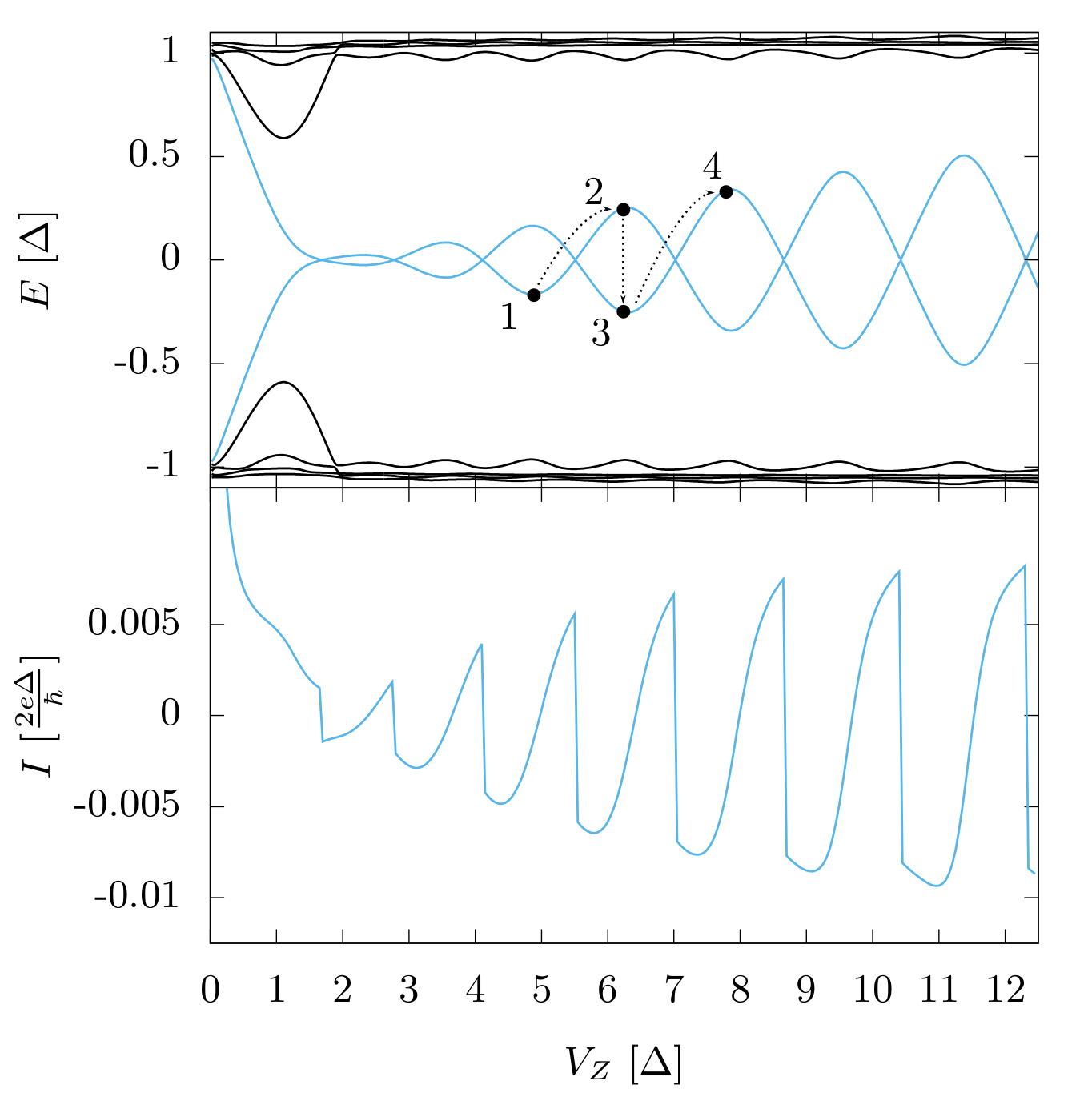}
	\caption{Spectrum of the Josephson junction at $\varphi=\frac{\pi}{2}$ (upper panel) with parameters as in Fig.~\ref{fig:effspec}c). The points and arrows visualize a measuring scheme to extract the contribution of the MBS to the Josephson current without the disguising background from higher energy states.
	First, an adiabatic sweep of the Zeeman field brings the initial state from point one to point two. Then the system will relax to the ground state again (visualized with point three) and the process can be repeated. The lower panel shows the resulting Josephson current at phase $\varphi=\frac{\pi}{2}$ as function of Zeeman field when the currents before and after the relaxtion are substracted.}
	\label{fig:Measure}
\end{figure}

Here, we want to propose an alternative experimental means to bring the contributions from the MBS to the critical current  to light. A sketch of this scheme is shown in Fig~\ref{fig:Measure}.
First, we consider a sweep of the magnetic field. At some point in parameter space there will be a crossing because of the protected parity in superconductors. When the sweep is done adiabatically
the parity will not change. The adiabaticity of the process gives a first time scale. However, on a larger second time scale the state will relax into the ground state due to quasi-particle poisoning. The Josephson current before and after this relaxation will include the same contribution from the background, 
but different contributions from the two distinct parity states of the non-local fermion built from the MBS. In fact, these MBS contributions differ in sign, so that the difference of the
Josephson current for a given phase difference before and after the relaxation to the ground state reveals only the MBS contributions (see Fig.~\ref{fig:Measure}).

%

\end{document}